\documentclass[unpublished]{JHEP3} 
\usepackage{latexsym}
\usepackage{epsfig}
\usepackage{amssymb}
\usepackage{amsthm}
\usepackage{amsmath}
\usepackage{amssymb,amsfonts}
\usepackage{verbatim}

\newcommand{\be}{\begin{eqnarray}}
\newcommand{\ee}{\end{eqnarray}}

\JHEPspecialurl{http://jhep.sissa.it/JOURNAL/JHEP3.tar.gz}

\usepackage{epsfig,multicol,bbm}

\newcommand\fverb{\setbox\fverbbox=\hbox\bgroup\verb}
\newcommand\fverbdo{\egroup\medskip\noindent%
            \fbox{\unhbox\fverbbox}\ }
\newcommand\fverbit{\egroup\item[\fbox{\unhbox\fverbbox}]}
\newbox\fverbbox

\title{P-nflation: generating cosmic Inflation with p-forms}

\author{Cristiano Germani\\
    LUTH, Observatoire de Paris, CNRS UMR 8102, Universit\'{e} Paris Diderot,
5 Place Jules Janssen, 92195 Meudon Cedex, France\\
    E-mail: \email{cristiano.germani@obspm.fr}}

\author{Alex Kehagias\\
    Department of Physics, National Technical University of Athens
GR-15773, Zografou, Athens, Greece\\
    E-mail: \email{kehagias@central.ntua.gr}}

\received{}       
\accepted{}       

\abstract{We show  that an inflationary background might be realized
by using any p-form non-minimally coupled
to gravity. Standard scalar field inflation corresponds to the 0-form case and vector inflation to the
1-form. Moreover, we show that
the 2- and 3-form  fields  are dual to a new vector and scalar inflationary theories where the kinetic terms are non-minimally coupled to gravity.}


\begin{document}


\section{Introduction}
Our Universe looks impressively smooth. At cosmological scales, the observable sky seems to be only a tiny fluctuation
out of a very homogeneous, isotropic and flat background, the Friedman-Robertson-Walker spacetime (FRW).
In fact, latest measures of the cosmic microwave background radiation (CMB) \cite{wmap}
clearly show that the CMB radiation is grossly homogeneous and isotropic. This implies that all points of the CMB sky
should have been in causal contact, with a high degree of accuracy, at very early times. In other words, two distant photons of the CMB
sky should have had enough time to meet each other and exchange information in the past. Unfortunately, General Relativity (GR) together with
the Standard Model of Particle Physics alone is not able to reproduce this simple fact. Standard Big Bang cosmology predicts a finite
``life time'' of the Universe which is too short to allow the observed high homogeneity and isotropy.
Moreover, an isotropic Universe is a highly unstable solution of GR sourced by standard matter. This is because, close to the Big Bang singularity,
the energy density associated with anisotropic perturbations is stronger that
the energy density of standard matter. An extra theoretical problem is the explanation of the observed spatial flatness of our Universe.
In standard cosmology, the ratio between the spatial curvature and the Hubble radius of the Universe
decreases monotonically backwards in time.
In this case, to have the required flatness today, an extreme
fine tuning of the spatial curvature must be used at very early times.

Many alternative ``solutions'' to these problems have been put forward in the past \cite{many}-\cite{bounce}. However, the most
developed and yet simple idea still remain inflation. Inflation solves the homogeneity, isotropy and flatness problems in one go just by postulating a
rapid expansion of the early time Universe post Big Bang. Nevertheless, a fundamental realization of this idea is still eluding us. Originally,
the effective theory of inflation has been realized by sourcing GR with a slow ``rolling'' massive scalar field \cite{chaotic} with minimal or even
non-minimal kinetic term \cite{k}.
Fundamental scalar fields are however not yet
discovered in nature so in principle the idea of inflation might well be realized  by other, more complex fields.
In this paper we explore the idea, initiated with
the vector inflation \cite{vec},\cite{vec1}, \cite{muk}, that inflationary backgrounds might alternatively obtained from p-forms
sourcing GR. Vector curvaton has been considered in \cite{dim}, vector anisotropies in an inflating background in \cite{soda}.
In particular we will show that any
p-form conveniently coupled with gravity might produce an inflationary background. In this view, the scalar and vector fields are only
the special 0- and 1-form cases of the general p-nflationary scenario.

Non-minimal coupling of fields with gravity
appear very often in many physical cases. For example, there are evidences that
a non minimal coupling of the Higgs field with gravity \cite{higgs}
might provide a very appealing particle physics scenario for inflation. Exotic fields such as higher spin fields can consistently
propagate only if non-minimally coupled
to gravity \cite{hs},\cite{hs1}. String theory compactifications
always introduce non minimal couplings  of geometric extra-dimensional fields and
four dimensional gravity \cite{hd}, \cite{hd1} and finally interesting successful models of Dark Energy
are based on non minimal couplings \cite{dark}.

Let us now see why non minimal coupling are fundamental to obtain an inflating background out of p-forms.
General p-forms minimally coupled to gravity cannot in fact support an inflationary background, not even in the case in which their gauge symmetry is explicitly
broken by a mass term. However, we will show here that there are always non-minimal interactions of massive p-forms with gravity that can support an
inflating background. The trick is to find the correct interactions reproducing the massive scalar field equations coupled with gravity out of p-forms. Let us
quickly discuss the 1-form case, vector inflation.

In a gravitational background, the field equations
\be
\nabla^\mu F_{\mu\nu}=m^2 A_\nu
\ee
for a massive vector field $A^\mu$ are written as
\be\label{example}
\square A_\mu-{R_\mu}^\nu A_\nu=m^2 A_\mu\ .
\ee
 We will explicitly work on a FRW background
\be
ds^2=-dt^2+a(t)^2 dx^i dx^j \delta_{ij}\ ,
\ee
and we will use Greek indices for four-dimensional quantities, $0$ for proper time and Latin indices for three-dimensional space.
By defining the re-scaled physical field \cite{muk,dim} $B_i\equiv \frac{A_i}{a}$ and taking into account the explicit form of $R_{ij}$ in terms of derivative
of the scale factor we have that (\ref{example}) becomes
\be\label{almost}
\ddot B_i+3H\dot B_i+(m^2-\frac{R}{6})B_i=0\ ,
\ee
where $H=\dot a/a$ is the Hubble constant and $\dot f=\frac{\partial f}{\partial t}$. Moreover, using the field equation for the temporal
component of the gauge field, one finds that $A_0=0$.

We note that the equation (\ref{almost}) almost reproduce a massive scalar field equation. To get rid of the un-wanted term proportional to the Ricci
scalar we can simply introduce in the massive U(1) action the following non-minimal term
\be\label{sufficient}
S_{nm}=\int d^4x \sqrt{-g}\frac{R}{12}A_\mu A^\mu\ .
\ee
This is the necessary non-minimal coupling defining the vector inflation of \cite{muk}.
Finally, by the symmetries of the background, the coupling (\ref{sufficient}) is also sufficient to reproduce
a scalar field type energy density driving an inflating background, {\it i.e.}
\be
\rho=-T^0{}_0=\frac{1}{2}\left(\dot B_i\dot B_i + m^2B_i B_i\right)\ ,
\ee
where summation over repeated indices is understood. The spatial Einstein equations in a FRW background
will however be incompatible with a vector field theory as it generically produces anisotropies which breaks the FRW symmetries. This problem is
nevertheless overcomes by considering 3 mutually orthogonal fields \cite{orth} or by a large number of randomly distributed fields \cite{muk}.

\section{P-nflation}

In this section we will explicitly consider massive non-minimally coupled p-form to gravity able to reproduce a slow roll inflationary scenario.

A massive p-form field $A_{\mu_1...\mu_p}$ in D-dimensions has
\be
\left(\begin{array}{c} D-1\\ p\end{array}\right) =\frac{(D\!-\!1)!}{(D\!-\!1\!-\!p)!p!}
\ee
degrees of freedom.
In particular, for $D=4$, the degrees of freedom of different forms are:
0-form 1 degree of freedom (the scalar field), 1-form 3 degrees of freedom (massive vector), 2-form 3 degrees of freedom and a 3-form has just 1 degree of freedom.
Note that, contrary to the 0- and 3-form fields which are consistent with
a homogeneous FRW cosmological model, we need a triplet of orthogonal \cite{orth}
 1- or 2-forms  (or, alternatively a large number of them \cite{muk}) to achieve homogeneity.
The cosmology of the 1-form (a massive U(1) gauge field) has been studied in \cite{muk}.
Here we will consider the remaining cases, namely, the 2- and the 3-form.

The general action able to reproduce a scalar field-like equation of motion of a FRW background, as explained in the introduction, is
\begin{eqnarray}\label{action}
\!\!\!&\!\!\!&\!\!\!\!S=\int d^4 x \sqrt{-g}\left\{\frac{1}{2\kappa^2} R-\frac{1}{2(p+1)!}F_{\alpha_1\ldots \alpha_{p+1}}F^{\alpha_1\ldots \alpha_{p+1}}-
\frac{1}{2p!}m^2 A_{\alpha_1\ldots\alpha_p}A^{\alpha_1\ldots\alpha_p}+\right.\nonumber \\&&
\left.+ \frac{p(2+3p-p^2)}{48}\, R\, A_{\alpha_1\ldots\alpha_p}
A^{\alpha_1\ldots\alpha_p}-
\frac{p(p-1)}{2\cdot p!}A_{\mu_1\ldots\mu_{p-1}\kappa}\,R^{\kappa\lambda}\, {A^{\mu_1\ldots \mu_{p-1}}}_\lambda\right\} \label{gener}\ ,
\end{eqnarray}
where
\be
F_{\mu_1\ldots \mu_{p+1}}=(p+1) \nabla_{[\mu_1} A_{\mu_2\ldots \mu_{p+1}]}\ .
\ee
In the next section, we will explicitly show that the action (\ref{gener}) reproduce inflationary backgrounds defining
what we shall call the p-nflationary scenarios.

In the following we will just consider the 2 and 3 forms as the 0 and 1 forms
can be found in the literature. In particular by taking $p=0$ we get the chaotic inflation of \cite{chaotic} and for $p=1$ we get the vector inflation of
\cite{muk}.

\subsection{The 2-form}

The action (\ref{gener}) for a 2-form field $A_{\mu\nu}$ is explicitly written as
\be\label{s2}
\!\!\!\!\!\!\!\!\!\!S_2=\int d^4 x \sqrt{-g} \left(\frac{1}{2\kappa^2} R-\frac{1}{12}F_{\mu\nu\rho}F^{\mu\nu\rho}-\frac{1}{4}m^2
A_{\mu\nu}A^{\mu\nu}+\frac{1}{6}R\, A_{\mu\nu}A^{\mu\nu}+\frac{1}{2}A_{\mu\kappa}\,R^{\kappa\lambda}\,{A_\lambda}^\mu\right)
\ee
where
\be
F_{\mu\nu\rho}=\nabla_\mu A_{\nu\rho}+\nabla_\rho A_{\mu\nu}+\nabla_\nu A_{\rho\mu}\ .
\ee

By varying the action (\ref{s2}) with respect to $A_{\mu\nu}$ we obtain the following field equations
\be
\nabla^\mu F_{\mu\nu\rho}-m^2 A_{\nu\rho}+\frac{2}{3} R\, A_{\nu\rho}+{R^\lambda}_{\nu}A_{\rho\lambda} -{R^\lambda}_{\rho}A_{\nu\lambda}=0\ .
\ee
We may express the above equations as
\be
\nabla^2 A_{\nu\rho}+\nabla_\rho\nabla^\mu A_{\mu\nu}-\nabla_\nu\nabla^\mu A_{\mu\rho}+m^2 A_{\nu\rho}+\frac{2}{3} R\, A_{\nu\rho}+
2{R^\lambda}_{\nu}A_{\rho\lambda} -2{R^\lambda}_{\rho}A_{\nu\lambda}=0 \label{B}\ .
\ee
The Einstein equations associated with (\ref{s2}) are
\be
R_{\mu\nu}-\frac{1}{2} g_{\mu\nu} R=\kappa^2T_{\mu\nu}\ ,
\ee
where
\be
T_{\mu\nu}&=&\frac{1}{2}F_{\mu\kappa\sigma}{F_\nu}^{\kappa\sigma}+(m^2\! -\!\frac{2}{3} R)
A_{\mu\kappa}{A_\nu}^{\kappa}-
A_{\mu\kappa}R^{\kappa\alpha}A_{\alpha\nu}-A_{\kappa\mu}{R_\nu}^\alpha {A_\alpha}^\kappa-
A_{\kappa\nu}{R_\mu}^\alpha {A_\alpha}^\kappa
\nonumber\\&&
-\frac{1}{2}\Big{[}\delta_\mu^\rho\nabla^\gamma\nabla_\rho +\delta_\nu^\rho \nabla^\gamma\nabla_\rho
-\delta_\mu^\rho\delta_\nu^\gamma \nabla^2
-g_{\mu\nu}\nabla^\rho\nabla_\gamma\Big{]}A_{\rho\kappa}{A^{\gamma}}^\kappa\nonumber\\ &&
-\frac{1}{3} R_{\mu\nu}A_{\alpha\beta}A^{\alpha\beta}-
\frac{1}{3} g_{\mu\nu}\nabla^2 A_{\alpha\beta}A^{\alpha\beta}
+\frac{1}{3} \nabla_\mu\nabla_\nu A_{\alpha\beta}A^{\alpha\beta}\nonumber\\ &&
+g_{\mu\nu}(-\frac{1}{12}F_{\kappa\lambda\rho}F^{\kappa\lambda\rho}-\frac{1}{4}m^2
A_{\kappa\lambda}A^{\kappa\lambda}+\frac{1}{6}R\, A_{\kappa\lambda}A^{\kappa\lambda}+
\frac{1}{2}A_{\alpha\kappa}\,R^{\kappa\lambda}\,{A_\lambda}^\alpha)\ .\label{t2}
\ee

Let us now consider an FRW geometry with metric
\be
ds^2=-dt^2+a(t)^2dx_idx^i \label{met}\ ,
\ee
and let us assume that the 2-form $A_{\mu\nu}$ is time-depended only. We may parameterize its components
as
\be
A_{0i}=a^2\,b_{0i}(t) ~,~~~ A_{ij}=a^2\,b_{ij}(t)\ .
\ee
Then the field equations (\ref{B}) are written as
\be
& & b_{0i}\left(2\dot{a}^2-m^2 a^2\right)=0\, , \\
&& \ddot{b}_{ij}+3\frac{\dot{a}}{a}\dot{b}_{ij}+m^2 b_{ij}=0\ .
\ee
Thus, we find that
\be
b_{0i}=0\ ,
\ee
and each one of the 3 components $b_{ij}$ satisfies equation identical to the inflaton field.

It is also straightforward to calculate the components of the energy-momentum tensor eq.(\ref{t2}).
We find that
\be
\!\!\!T_{00}&=&\frac{1}{4}(\dot{b}_{mn}\dot{b}_{mn}+m^2b_{mn}b_{mn})\ ,
\\
\!\!\!T_{ij}&=&\left\{\frac{1}{12} b_{mn}b_{mn}\left(16\frac{\dot{a}^2}{a^2}+8\frac{\ddot{a}}{a}-3 m^2\right)
+\frac{10}{3}\frac{\dot{a}}{a}\dot{b}_{mn}b_{mn}+\frac{11}{12}\dot{b}_{mn}\dot{b}_{mn}+
\frac{2}{3}\ddot{b}_{mn}b_{mn}\right\}
g_{ij}+\nonumber\\&&\!\!\!\!\!\!\!\!\!\!\!\!\!\!\!\!\!\!\!\!\!\!
+b_{ik}b_{jk}(4\dot{a}^2+6 a \ddot{a}-m^2 a^2)-\frac{1}{2} a^2(\ddot{b}_{ik}b_{jk}+b_{ik}\ddot{b}_{jk})+2a^2\dot{b}_{ik}
\dot{b}_{jk}+\frac{7}{2}a\dot{a}(b_{ik}\dot{b}_{jk}+\dot{b}_{ik}b_{jk})\ ,\label{tij}
\ee
where summation over repeated indices is understood. The spatial part of the energy momentum tensor can however be simplified by using the
time component of the Bianchi identities ($\nabla_\alpha T^\alpha{}_\beta=0$) and the field equations
to substitute for the $\ddot a$ and $\ddot b_i$ terms into (\ref{tij}). We then obtain
\be
T^i_j&=&\left[\left(\frac{3}{8}m^2-\frac{H^2}{3}\right)b_{lm}b_{lm}-\frac{2}{3}Hb_{lm}\dot b_{lm}-\frac{11}{24}\dot b_{lm}\dot b_{lm}\right]\delta^i_j+\cr
&+&H^2 b_{ik}b_{jk}+H(\dot b_{ik}b_{jk}+b_{ik}\dot b_{jk})+\dot b_{ik}\dot b_{jk}-\frac{3}{4}m^2 b_{ik}b_{jk}\ .
\ee
As we shall see later, the 2-form inflation is dual to a vector inflation theory that differs from the one of \cite{muk}. However we will now show that at the
background level the two theories behave exactly the same. Let us spatially dualize $b_{ij}$ as follows:
\be
b_{ij}=\varepsilon_{ijk}B^k\ ,
\ee
where $\varepsilon_{ijk}$ is the totally antisymmetric symbol. We then get
\be
T_{00}=\frac{1}{2}\left(\dot B_i\dot B_i+ m^2 B_i B_i\right)\ ,
\ee
and
\be
T^i_j=\left(\frac{H^2}{3}B_i B_i+\frac{2H}{3}B_i \dot B_i+\frac{1}{12}\dot B_i \dot B_i\right)\delta^i_j-
H^2B_i B_j &-&H\left(\dot B_i B_j+B_i \dot B_j\right)-\cr &-&\dot B_i \dot B_j+\frac{3}{4}m^2 B_i B_j\ ,
\ee
whereas the equation of motions are still scalar field-like
\be
\ddot B_i+3H\dot B_i+m^2 B_i=0\ .
\ee
The stress tensor we just found is identical to the one of vector inflation \footnote{Note that in \cite{muk} Bianchi identities and equation of motion
were not used to write down $T_{ij}$. By employing Bianchi identities and equation of motion the stress tensor of \cite{muk} indeed simplifies to ours.}.
It is then clear that, as happens for vector inflation, off diagonal components of the stress tensor do not satisfy Einstein
equations with a FRW background. However, this problem can be circumvented by either
taking a large number of fields randomly distributed or by considering three mutually orthogonal
vector fields $B_i^{(a)}=\delta^{(a)}_iB$ where $(a=1,2,3)$ and $B$ a single scalar
so that $\sum_{(a)} B_i^{(a)} B_j^{(a)}=\delta^i_j B^2$ \cite{orth}. Note that in terms of the antisymmetric 2-form field, this corresponds
to $b_{ij}^{(a)}={\varepsilon^{a}}_{ij}B$.

In the second case, the 2-nflationary theory in a FRW background reduces to the massive scalar field theory \cite{chaotic}, in which $B$ is identified
as the scalar field. Explicitly we obtain the inflationary closed system
\be
\ddot B+3H\dot B+m^2 B=0\ ,\cr
H^2=\frac{4\pi G}{3}\left(\dot B^2+m^2 B^2\right)\ .
\ee

In the case of a large number $N$ of randomly distributed fields we instead have that \cite{muk}
\be
\sum_{(a)} B_i^{(a)} B_j^{(a)}\simeq \frac{N}{3}\delta^i_j B^2+O(1) \sqrt{N}B_i B_j\ ,
\ee
so in this case sub-leading anisotropies survives after inflation as explained in \cite{muk}. Usual inflation is then defined in slow roll formalism. We would also
like to mention that a more general model can be obtained by a more general potential $V(A_{\mu\nu}A^{\mu\nu})$ instead of the simplest
potential $\frac{1}{2}m^2A_{\mu\nu}A^{\mu\nu}$ here considered.

\subsection{The 3-form}

We will now turn our attention to the 3-form field $A_{\mu\nu\rho}$ case. Its action is explicitly
\begin{eqnarray}\label{S3}
S_3=\int d^4 x \sqrt{-g} \{\frac{1}{2\kappa^2} R&-&\frac{1}{48}F_{\mu\nu\rho\sigma}F^{\mu\nu\rho\sigma}-\frac{1}{12}m^2
A_{\mu\nu\rho}A^{\mu\nu\rho}+\frac{1}{8}R\, A_{\mu\nu\rho}A^{\mu\nu\rho}-\cr
&-&\frac{1}{2}A_{\mu\nu\kappa}\,R^{\kappa\lambda}\,{A_\lambda}^{\mu\nu}\}
\label{action3}
\end{eqnarray}
where
\be
F_{\mu\nu\rho\sigma}=\nabla_\mu A_{\nu\rho\sigma}-\nabla_\sigma A_{\mu\nu\rho}+\nabla_\rho A_{\sigma\mu\nu}-\nabla_\nu A_{\rho\sigma\mu}\ .
\ee
The field equations for the 3-form is
\be
\nabla_\kappa F^{\kappa\mu\nu\rho}-m^2 A_{\mu\nu\rho}+\frac{3}{2} R\, A_{\mu\nu\rho}-2(
{R^\kappa}_{\rho}
A_{\mu\nu\kappa}+{R^\kappa}_{\nu}
A_{\rho\mu\kappa}+{R^\kappa}_{\mu}
A_{\nu\rho\kappa})=0\ ,
\ee
which may be written as
\be
&&\nabla^2 A_{\mu\nu\rho}-\nabla_\sigma\nabla^\mu A_{\mu\nu\rho}+\nabla_\rho\nabla^\mu A_{\mu\nu\sigma}-
\nabla_\nu\nabla^\mu A_{\mu\rho\sigma}-m^2 A_{\mu\nu\rho}\nonumber \\&&+
{R_{\mu\nu}}^{\lambda\sigma}A_{\rho\lambda\sigma}+{R_{\rho\mu}}^{\lambda\sigma}A_{\nu\lambda\sigma}+
{R_{\nu\rho}}^{\lambda\sigma}A_{\mu\lambda\sigma}\nonumber\\
&&-3( {R^\kappa}_{\rho} A_{\mu\nu\kappa}+{R^\kappa}_{\nu}
A_{\rho\mu\kappa}+{R^\kappa}_{\mu} A_{\nu\rho\kappa})=0\, .
\label{a3}
\ee
The Einstein equations take the standard form
\be
R_{\mu\nu}-\frac{1}{2} g_{\mu\nu} R=\kappa^2T_{\mu\nu}\ ,
\ee
where
\be
T_{\mu\nu}&=&\frac{1}{6}F_{\mu\kappa\sigma\rho}{F_\nu}^{\kappa\sigma\rho}+(\frac{1}{2}m^2\! -\!\frac{3}{4} R)
A_{\mu\kappa\sigma}{A_\nu}^{\kappa\sigma}+A_{\alpha\beta\gamma}{R^\gamma}_\mu {A_{\nu}}^{\alpha\beta}+
A_{\alpha\beta\mu} R_{\gamma\nu} A^{\gamma\alpha\beta}\nonumber\\&&
+A_{\alpha\mu\gamma}R^{\gamma\beta}{{A_\beta}^\alpha}_\nu
+A_{\mu\beta\gamma}R^{\gamma \sigma}{A_{\sigma\nu}}^\beta
-\frac{1}{2}\Big{[}(\delta_\mu^\rho\delta_\nu^\gamma+\delta_\nu^\rho\delta_\mu^\gamma)
\nabla_\beta\nabla_\rho
-\delta_\mu^\gamma g_{\beta\nu}\nabla^2\nonumber\\&&
-g_{\mu\nu}\nabla_\beta\nabla_\gamma\Big{]}A_{\kappa\lambda\gamma}A^{\beta\kappa\lambda}
-\frac{1}{4} R_{\mu\nu}A_{\alpha\beta\gamma}A^{\alpha\beta\gamma}-
\frac{1}{4} g_{\mu\nu}\nabla^2 A_{\alpha\beta\gamma}A^{\alpha\beta\gamma}
+\frac{1}{4} \nabla_\mu\nabla_\nu A_{\alpha\beta\gamma}A^{\alpha\beta\gamma}\nonumber\\ &&
+g_{\mu\nu}(-\frac{1}{48}F_{\kappa\lambda\rho\sigma}F^{\kappa\lambda\rho\sigma}-\frac{1}{12}m^2
A_{\kappa\lambda\rho}A^{\kappa\lambda\rho}+\frac{1}{8}R\, A_{\kappa\lambda\rho}A^{\kappa\lambda\rho}-
\frac{1}{2}A_{\kappa\lambda\sigma}\,R^{\sigma\rho}\,{A_\rho}^{\kappa\lambda})\label{tmn3}\ .
\ee
Again for an FRW background, of the form (\ref{met}), we may assume
a time-depended 3-form. The components of this form may be
parameterized in terms of a 2-form $a_{ij}$ and a scalar $\phi$ as
\be A_{0ij}=a_{ij}(t) \, , ~~~~~A_{ijk}=\phi(t) \epsilon_{ikj}\ ,
\label{a33} \ee
where $\epsilon_{ijk}$ is the spatial volume
element.
Then the field equations (\ref{a3}) are explicitly written
as \be
&& a_{ij}\left(\frac{\ddot{a}}{a}-\frac{\dot{a}^2}{a^2}+m^2\right)=0\, , \\
&& \ddot{\phi}+3\frac{\dot{a}}{a}\dot{\phi}+m^2\phi=0 \label{phi}\ .
\ee
Thus, for an expanding solution, we have
\be
a_{ij}=0\ ,
\ee
and the single component of a massive 3-form non-minimally coupled to gravity parameterized by the scalar
$\phi$ satisfies the equation of a potential inflaton.

Plugging (\ref{a33}) in the energy-momentum tensor (\ref{tmn3}), we find that
\be
&& T_{00}=\frac{1}{2}\dot{\phi}^2+\frac{1}{2}m^2 \phi^2\, , ~~~T_{0i}=0\nonumber \\&&
T_{ij}=a^2\delta_{ij}\left(3\frac{\dot{a}}{a}\dot{\phi}\phi +\frac{1}{2}m^2 \phi^2+\frac{1}{2} \dot{\phi}^2+\phi\ddot{\phi}\right)\ .
\ee
Using (\ref{phi}) in the Einstein equations we get
\be
&&3 \frac{\dot{a}^2}{a^2}=\frac{1}{2}\dot{\phi}^2+\frac{1}{2}m^2 \phi^2\ ,\\
&& 2\frac{\ddot{a}}{a}+\frac{\dot{a}^2}{a^2}=-\frac{1}{2}\dot{\phi}^2+\frac{1}{2}m^2 \phi^2\ ,
\ee
which are the standard equations for a scalar field. Thus, the action for a 3-form non-minimally coupled to gravity
describes a standard scalar field minimally coupled to gravity, at least at the background level. In this case the system is therefore
automatically compatible with isotropy of the background.

\section{Dual Theories}
As we have already noticed, in four spacetime dimensions, 3 and 4 forms are  dual to vector and scalar fields, respectively. One may therefore
wonder whether the 2-3 -nflation theories are just the scalar and vector field inflationary scenarios in their dual description.
We will show in the following that this is not the case and 2- and 3-nflation are completely new theories with
no relation to scalar and vector inflation. For comparison we will however call these new theories dual-vector and dual-scalar field inflations.
In fact, as we shall see it,
the dual 2- and 3-nflation theories are higher derivatives theories. The vector and scalar fields are indeed
coupled with inverse of curvatures in p-nflation.

\subsection{Dual Vector inflation}

The action (\ref{s2}) may be equivalently written as
\be
S_2=\int d^4 x \sqrt{-g} \left(\frac{1}{2\kappa^2} R+\frac{1}{12}F_{\mu\nu\rho}F^{\mu\nu\rho}+
\frac{1}{2}A_{\mu\nu}\nabla_\sigma F^{\sigma\mu\nu}+\frac{1}{2}A_{\mu\kappa}\,
M^{\kappa\lambda}\,{A_\lambda}^{\mu}\right) \label{action2}\ ,
\ee
where
\be
M^{\kappa\lambda}= g^{\kappa\lambda}\left(\frac{R}{3}-\frac{m^2}{2}\right)+R^{\kappa\lambda}\ .
\ee
Integrating out $F_{\mu\nu\rho}$ we get the original action (\ref{s2}).

We now define the dual fields
\be
F_{\mu\nu\rho}=m\ \epsilon_{\mu\nu\rho\sigma}A^\sigma\ ,\cr
A_{\alpha\beta}=\epsilon_{\alpha\beta\mu\nu}B^{\mu\nu}\ ,
\ee
where $\epsilon_{\mu\nu\rho\sigma}$ is the spacetime volume element.

The dual action then reads
\be\label{dual}
{\cal L}_{dual}=\frac{1}{2\kappa^2}R-\frac{1}{2}m^2 A_\mu A^\mu -m B^{\alpha\beta}F_{\alpha\beta}+  m^2 B^{\alpha\lambda}B_{\alpha \beta}\tilde\Delta^{\beta \lambda}\ ,
\ee
where
\be
\tilde\Delta^{\alpha\beta}=\left(1-\frac{5}{3}\frac{R}{m^2}\right)\delta^{\alpha\beta}+2\frac{R^{\alpha\beta}}{m^2}\ ,
\ee
and $F_{\mu\nu}=\nabla_\mu A_\nu-\nabla_\nu A_\mu$ as usual.
We now define
\be
\Delta^{\mu\nu}_{\alpha\beta}\equiv \delta^{\mu\lambda}_{\alpha\beta}\tilde\Delta^\nu{}_\lambda\ ,
\ee
and the ``inverse'' tensor
\be
\Lambda^{\alpha\beta}_{\gamma\delta}\Delta^{\mu\nu}_{\alpha\beta}=\delta^{\mu\nu}_{\gamma\delta}\ ,
\ee
where $\delta^{\mu\nu}_{\alpha\beta}=2\delta^{\mu}{}_{[\alpha} \delta_{\beta]}{}^{\nu}$ is the generalized Kroneker delta.

With the above definitions we can integrate out $B_{\alpha\beta}$ from (\ref{dual}) and obtain the dual Vector inflation
\be
{\cal L}_{dual}=\frac{1}{2\kappa^2}R-\frac{1}{4}\Lambda^{\mu\nu}{}_{\rho\sigma}F_{\mu\nu} F^{\rho\sigma} -\frac{1}{2}m^2 A_\mu A^\mu\ . \label{dual2}
\ee
The action we just found clearly differs from the vector inflation action of \cite{muk}. It is
however interesting to see that in a flat background the above action
just reduces to the massive $U(1)$ gauge field
\be
{\cal L}_{dual}=-\frac{1}{4}F_{\mu\nu} F^{\mu\nu}-\frac{1}{2}m^2 A^2\ .
\ee

We should stress here that the basic difference between the action  (\ref{dual2}) and the corresponding action for the massive vector field of
\cite{muk} is that in \cite{muk} the vector field has normal kinetic term and a non-minimal mass term whereas, in our case,
the dual vector field has normal mass term but non-minimal kinetic term.

\subsection{Dual Scalar inflation}

We finally turn our attention to the three-form action. The action (\ref{S3}) may be equivalently written as
\be
S_3=\int d^4 x \sqrt{-g} \left(\frac{1}{2\kappa^2} R+\frac{1}{48}F_{\mu\nu\rho\sigma}F^{\mu\nu\rho\sigma}+
\frac{1}{6}A_{\mu\nu\rho}\nabla_\sigma F^{\mu\nu\rho\sigma}-\frac{1}{2}A_{\mu\nu\kappa}\,
M^{\kappa\lambda}\,{A_\lambda}^{\mu\nu}\right) \label{action323}\ ,
\ee
where
\be
M^{\kappa\lambda}= g^{\kappa\lambda}\left(\frac{m^2}{6}-\frac{R}{4}\right)+R^{\kappa\lambda}\ .
\ee
Integrating out $F_{\mu\nu\rho\sigma}$ we get back the original action (\ref{S3}).
The dual theory is now obtained by expressing the field strength and the three-form potential in dual fields, {\it i.e.},
\be
F_{\mu\nu\rho\sigma}=m \epsilon_{\mu\nu\rho\sigma} \Phi\ ,
\ee
and
\be
A_{\mu\nu\rho}=\epsilon_{\mu\nu\rho\alpha}B^\alpha\ .
\ee
The dual action then reads
\be
{\cal{L}}_{dual}=\frac{1}{2\kappa^2}R-\frac{1}{2}m^2\Phi^2-m B^\alpha\partial_\alpha\Phi+\frac{m^2}{2}\Delta^{\alpha\beta}
B_{\alpha} B_{\beta}\ , \label{dual3}
\ee
with
\be
\Delta^{\alpha\beta}=\left(1+\frac{R}{2m^2}\right)g^{\alpha\beta}-\frac{2}{m^2}R^{\alpha\beta}\ .
\ee
The effective theory for the scalar field $\Phi$ is then obtained by integrating out $B_{\alpha}$. By defining
\be
\Lambda^{\alpha\mu}\Delta_{\alpha\nu}=\delta^\mu_\nu\ ,
\ee
we have
\be
{\cal{L}}_{dual}=\frac{1}{2\kappa^2}R-\frac{1}{2}\Lambda^{\alpha\beta}\partial_\alpha\Phi\partial_\beta\Phi-\frac{1}{2}m^2\Phi^2\ .
\ee
The dual theory of the 3-nflation is therefore a higher curvature theory and so it
differs from the standard scalar field
inflation in which $\Lambda^{\alpha\beta}=g^{\alpha\beta}$.

It is interesting to see that the Einstein equations in the dual formulation looks much simpler than in the three-form form.
Variation of the above action with respect the metric produce in fact the following Einstein equations
\be
R_{\mu\nu}-\frac{1}{2}g_{\mu\nu}R=\kappa^2 T_{\mu\nu}
\ee
where
\begin{eqnarray}
T_{\mu\nu}&=& \left(1+\frac{R}{2m^2}\right)\xi_\mu\xi_\nu-\frac{\xi^2}{2m^2}R_{\mu\nu}-\frac{1}{2}g_{\mu\nu}
\left(m^2\Phi^2+\xi^\kappa\partial_\kappa \Phi\right)\nonumber \\
&&-\frac{1}{m^2}\left(\nabla^\kappa\nabla_\mu S_{\nu\kappa}+\nabla^\kappa\nabla_\nu S_{\mu\kappa}-g_{\mu\nu}
\nabla^\kappa\nabla^\lambda S_{\kappa\lambda}-\Box S_{\mu\nu}\right)\ ,
\end{eqnarray}
with
\be
\xi^\mu=\Lambda^{\mu\nu} \partial_\nu\Phi\, , ~~~~~
S^{\mu\nu}=\xi^\mu\xi^\nu -\frac{1}{4}g_{\mu\nu}\xi^2\ .
\ee
In addition, the scalar field equation may be written
\be
\nabla_\mu\xi^\mu=m^2 \Phi\ .
\ee
It is very interesting to notice that for a deSitter background, in which $a(t)=e^{Ht}$, we have that
$\Lambda^{\alpha\beta}=g^{\alpha\beta}$. The dual action in a deSitter background therefore
reduces exactly to the minimally coupled scalar field action. This is not of course the case for any other
background. For example we expect that cosmological perturbations in these theory will greatly differ from the chaotic inflationary case. Although the check of the
cosmological perturbation for an inflating background is of primary interest we will however leave this for future work.

We also note that, as in the dual-vector case the basic difference between the action  (\ref{dual3}) and the standard chaotic inflation is that
in the case of chaotic inflation, the inflaton field has both normal kinetic and mass term, whereas in the dual-scalar theory, the dual to the
3-form scalar field has normal mass term but non-minimal kinetic term.

\section{Stability}

It has been claimed in \cite{hcp} that vector inflation is
probably unstable. This is due to a possible ghost instability of the longitudinal component of the massive U(1) vector in the
slow-roll regime. Such instability, may more easily be discussed after restoring gauge invariance by means of Stuckelberg fields. Let us
consider the following lagrangian
\be
{\cal{L}}_1=\frac{1}{2 \kappa^2} R-\frac{1}{4} F^2-\left[\frac{1}{2}m^2-\frac{1}{6}
R\right]\Big{(}\partial_\mu\chi -A_\mu\Big{)}\Big{(}\partial^\mu
\chi-A^\mu\Big{)}\ ,\label{Stu}
\ee
invariant under the following gauge transformations
\be
\delta A_\mu=\partial_\mu \theta\,, ~~~~~\delta\chi=\theta\ .
\ee
The original vector inflation action of \cite{muk} can be obtained by gauge fixing the lagrangian (\ref{Stu}) with
$\chi=\rm{const.}$.
The Stuckelberg field is clearly a ghost for
\be
\frac{1}{2}m^2-\frac{1}{6} R<0\ .
\ee
In slow roll ($\dot H\ll m^2\ll H^2$) and if no gravitational perturbations are considered, we have that
\be
\frac{1}{2}m^2-\frac{1}{6} R=\frac{1}{2}m^2-\dot{H}-H^2<0\ ,
\ee
so the field part of the theory (\ref{Stu}) on a fixed slow rolling
background suffers from a ghost instability. However, in order to finally decide whether or not vector
inflation is perturbatively unstable, the full gravitational and field
theory perturbation
analysis must be performed. This important task is left for future work.

Similarly, for the other p-nflationary theories, one may perform the same qualitative analysis by introducing appropriate
Stuckelberg form fields to restore gauge invariance.

For the 2-form inflation, we have
\be
{\cal{L}}_2=\frac{1}{2\kappa^2}
R-\frac{1}{12}F^2-\left[\left(\frac{1}{4}m^2-\frac{1}{6}R\right)\delta_\kappa^\rho+
\frac{1}{2}R_\kappa^\rho\right]\Big{(}\Theta_{\mu\rho}-B_{\mu\rho}\Big{)}\Big{(}\Theta^{\mu\kappa}
-B^{\mu\kappa}\Big{)}\ee
where
\be
\Theta_{\mu\nu}=\partial_\mu \chi_\nu-\partial_\nu \chi_\mu\ .
\ee
The above Lagrangian is invariant under
\be
\delta B_{\mu\nu}=\partial_\mu \theta_\nu-\partial_\nu
\theta_\mu\,, ~~~~~\delta\chi_\mu=\theta_\mu\ .
\ee
Again the Stuckelberg field $\chi_\mu$ behaves as a ghost if the
matrix
\be(M^2)_\kappa^\rho=\left(\frac{1}{4}m^2-\frac{1}{6}R\right)\delta_\kappa^\rho+\frac{1}{2}R_\kappa^\rho\ ,\ee
has negative eigenvalues. The eigenvalues of this matrix turns out to
be
\be(M^2)_0^0=\frac{1}{4}(m^2+2\dot{H}-2H^2)\,,
~~~~(M^2)_i^j=\frac{1}{4}(m^2-2\dot{H}-2H^2)\delta_i^j\ee
which are again negative for slow roll. Thus, the Stuckelberg field might produce a ghost instability.

Finally, for the 3-form inflation, we have
\begin{eqnarray}
\!\!\!\!\!\!\!\!\!\! {\cal{L}}_3=\frac{1}{2\kappa^2}
R-\frac{1}{48}F^2-\left[\left(\frac{1}{12}m^2-\frac{1}{8}R\right)\delta_\kappa^\rho+
\frac{1}{2}R_\kappa^\rho\right]\Big{(}\Theta_{\mu\sigma\rho}-A_{\mu\sigma\rho}\Big{)}\Big{(}\Theta^{\mu\sigma\kappa}
-A^{\mu\sigma\kappa}\Big{)}\ ,\end{eqnarray}
where
\be
\Theta_{\mu\nu\kappa}=\partial_\mu
\chi_{\nu\kappa}+\partial_\kappa
\chi_{\mu\nu}+\partial_\nu\chi_{\kappa\mu}
\ee
and $\chi_{\mu\nu}=-\chi_{\nu\mu}$.
The Lagrangian is now invariant
under
\be\delta A_{\mu\nu\kappa}=\partial_\mu
\theta_{\nu\kappa}+\partial_\kappa \theta_{\mu\nu}+\partial_{\nu}
\theta_{\kappa\mu}\,,
~~~~~\delta\chi_{\mu\nu}=\theta_{\mu\nu}\ .\ee
Again, the Stuckelberg field $\chi_{\mu\nu}$ would be a ghost if the
matrix
\be
(M^2)_\kappa^\rho=\left(\frac{1}{12}m^2-\frac{1}{8}R\right)\delta_\kappa^\rho+\frac{1}{2}R_\kappa^\rho \label{mat}\ee
has negative eigenvalues. However, in this case the matrix (\ref{mat}) is diagonal with eigenvalues
\be
(M^2)_0^0=\frac{1}{12}(m^2+9\dot{H})\,,
~~~~(M^2)_i^j=\frac{1}{12}(m^2-3\dot{H})\delta_i^j
\ee
which are positive in the slow roll regime.
Thus, the Stuckelberg field is in this case a canonical field and no ghost instability might be produced.

\section{Conclusions}
We have explored here the idea that inflationary backgrounds might
be  obtained from p-forms sourcing GR. This is in a sense a
generalization of  vector inflation \cite{vec},\cite{vec1},
\cite{muk} for higher order antisymmetric fields.  In particular, we have
shown that any p-form conveniently coupled to gravity might
produce an inflationary background. In this respect, the scalar
and vector fields are only the special 0- and 1-form cases of the
general p-nflationary scenario.

The basic ingredient for p-nflation is to introduce a non-minimal coupling of the massive p-form fields to gravity able to mimic
a slow rolling inflationary background. Moreover, for 1- and 2-nflation, a triplet of fields appropriately oriented or a large number of
them randomly distributed  are need in order to have configurations compatible with a homogeneous FRW background. For the standard chaotic
inflation (0-nflation) and 3-nflation, this is not
necessary, as scalars and 3-forms are compatible with homogeneity.

One may think that, due to Poincare duality which connects in general
 p-forms in $n$ dimensions with $n-p$-forms,  the 3-nflation will be identical
with chaotic inflation and 2-nflation with the vector one. However, as we have shown,
 the massive vector field in vector inflation has normal kinetic term and a non-minimal mass term, whereas the dual vector of the 2-nflation has normal
 mass term but non-minimal kinetic term.
Similarly, the inflaton scalar field has both normal kinetic and mass term, whereas  the dual-scalar theory of 3-nflation has normal mass term but
non-minimal kinetic term.

Finally, we would like to mention that, in p-nflation, scalar and tensorial
perturbations will generically differ (at slow roll order) from the standard perturbations generated in
chaotic inflation (and similar standard inflationary theories). This is due to the new algebraic structure of the p-nflationary theories.
For example, the perturbed pressures related to the p-forms will generically be anisotropic.
We therefore believe that
p-nflation would potentially lead to striking signatures in the CMB that can be in principle observed in
future experiments. The perturbation analysis
is however too involved to be performed in this paper as standard techniques must be revisited, we therefore leave this important task
for a subsequent work.

\acknowledgments

CG wishes to thank SISSA for partially supporting this work. CG also wishes to thank
Viatcheslav Mukhanov for useful discussions during the development of this work and
Lotfi Boubekeur and Fernando Quevedo for important comments on the dual theories.
AK wishes to thank support from the PEVE-NTUA-2007/10079 programme. This work is partially supported by the European
Research and Training Network MRTPN-CT-2006 035863-1.

\end{document}